# Magnetohydrodynamic Slow Mode with Drifting He++: Implications for Coronal Seismology and the Solar Wind

Joseph V. Hollweg

Space Science Center

Morse Hall

University of New Hampshire

Durham, NH 03824, USA

joe.hollweg@unh.edu

Daniel Verscharen

Space Science Center

Morse Hall

University of New Hampshire

Durham, NH 03824, USA

daniel.verscharen@unh.edu

Benjamin D.G. Chandran

Space Science Center and Department of Physics

Morse Hall

University of New Hampshire

Durham, NH 03824, USA

benjamin.chandran@unh.edu

Short Title: Oblique slow mode with drifting He++

Corresponding Author: Joseph V. Hollweg

joe.hollweg@unh.edu

## ABSTRACT

The MHD slow mode wave has application to coronal seismology, MHD turbulence, and the solar wind where it can be produced by parametric instabilities. We consider analytically how a drifting ion species (e.g. He$^{++}$) affects the linear slow mode wave in a mainly electron-proton plasma, with potential consequences for the aforementioned applications. Our main conclusions are: 1. For wavevectors highly oblique to the magnetic field, we find solutions that are characterized by very small perturbations of total pressure. Thus, our results may help to distinguish the MHD slow mode from kinetic Alfvén waves and non-propagating pressure-balanced structures, which can also have very small total pressure perturbations. 2. For small ion concentrations, there are solutions that are similar to the usual slow mode in an electron-proton plasma, and solutions that are dominated by the drifting ions, but for small drifts the wave modes cannot be simply characterized. 3. Even with zero ion drift, the standard dispersion relation for the highly oblique slow mode cannot be used with the Alfvén speed computed using the summed proton and ion densities, and with the sound speed computed from the summed pressures and densities of all species. 4. The ions can drive a non-resonant instability under certain circumstances. For low plasma beta, the threshold drift can be less than that required to destabilize electromagnetic modes, but damping from the Landau resonance can eliminate this instability altogether, unless $T_e / T_p >> 1$.

# 1. Introduction

Magnetohydrodynamic (MHD) waves have long been thought to play a role in heating the solar chromosphere and corona (e.g. Osterbrock 1961) and heating and accelerating the solar wind (e.g. the historical review by Hollweg 2008). MHD waves may dissipate via a turbulent cascade (e.g. Bruno & Carbone 2013; Tu & Marsch 1995) and there is growing evidence that the solar wind and the regions of the corona from which it originates may indeed be heated by MHD turbulence (e.g. Chandran & Hollweg 2009; Cranmer 2009; Cranmer et al. 2007; Ofman 2005; Verdini & Velli 2007; Verdini et al. 2010, 2012). Moreover, the turbulence may be the source of high-frequency waves which are cyclotron-resonant with protons and heavy ions (especially fully-ionized helium, i.e. $\alpha$-particles, which are the most abundant heavy ion) and which may explain why heavy ions in the fast solar wind are both hotter and flowing faster than the protons (Hollweg 2006, 2008; Hollweg & Isenberg 2002; Isenberg & Vasquez 2007, 2009, 2011; Kasper et al. 2013; Verdini & Velli 2007). On the other hand, an explanation is required for how heavy ions get decelerated from their high speeds in interplanetary space close to the Sun to the lower speeds observed at greater distances, e.g. at 1AU (Hellinger & Trávníček 2006; Kaghashvili et al. 2003; Li & Li 2008; Lu et al. 2006, Verscharen & Chandran 2013).

There are three MHD wave modes: the fast, slow and Alfvén (or intermediate) modes, but the Alfvén mode has received by far the greatest attention in the solar context, both because it is observed to be a principal component of the solar wind, and because it can carry substantial energy fluxes in virtue of its large group velocity. The ion-cyclotron wave is the high-frequency extension of the Alfvén mode, and it too has played a large role in considerations of coronal and solar wind heating and acceleration (e.g. Hollweg & Isenberg 2002; Kasper et al. 2013), and there may be some direct evidence for its presence in the solar wind (e.g. He et al. 2011); a full discussion of high-frequency waves and their roles in the corona and solar wind is beyond the scope of this paper.

From direct *in situ* spacecraft observations we have known for over four decades that Alfvén waves are a major component of fluctuations in the solar wind (Belcher et al. 1969; Belcher & Davis 1971; Coleman 1966, 1967, 1968; Unti & Neugebauer 1968). Particularly in the fast wind, these waves are mainly propagating outward from the Sun, suggesting, but not proving, a solar source. It was immediately realized that if the waves do indeed originate at the Sun with sufficient energy flux, they might account for the chromospheric and coronal heating, and drive the solar wind. Recent years have seen growing evidence that Alfvén waves are indeed present much closer to the Sun than the regions explored by spacecraft; see the recent reviews by De Moortel & Nakariakov (2012) and Banerjee et al. (2011). Ulrich (1996) presented spectroscopic observations suggesting the presence of Alfvén waves in a high magnetic field region in the chromosphere, with an upward Poynting flux sufficient to power coronal active regions. In newer observations the presence of Alfvénic waves in the chromosphere, transition region, and corona with periods of several minutes has been confirmed by direct imaging of structures which appear to be moving or swaying in the manner expected for Alfvénic waves (De Pontieu et al. 2007; McIntosh et al. 2011; Okamoto et al. 2007; Tomczyk et al. 2007; Tomczyk & McIntosh 2009). Recent spectroscopic observations also suggest the presence of Alfvén waves in the chromosphere and low corona (Banerjee et al. 2009; Gupta et al. 2010; Jess et al. 2009; Singh et al. 2011; see also Hassler et al. 1990); Bemporad & Abbo (2012), using spectroscopic data from EIS on Hinode, even provide evidence for significant energy deposition by low-frequency Alfvén waves in a polar coronal hole. At greater heights, in the region where the solar wind is believed to undergo its principal acceleration, Hollweg et al. (1982, 2010) looked at Faraday rotation fluctuations of radio signals from the Helios spacecraft during their superior conjunctions with the Sun. It was argued that the source of the observed fluctuations was most likely Alfvén waves associated with the waves observed by spacecraft much further from the Sun. The data were found to agree well with models (Chandran & Hollweg 2009; Cranmer & van Ballegooijen 2005; Cranmer et al. 2007; Verdini & Velli 2007; Verdini et al. 2010, 2012) in which turbulently damped Alfvén waves of solar origin heat and accelerate the solar wind. Thus, the Alfvén wave is a serious candidate for carrying the energy which heats and accelerates the solar wind and corona.

In addition to providing theoretical explanations for the heating and acceleration of the solar atmosphere, MHD waves are now being used as an observational tool for inferring the solar atmosphere's properties. This burgeoning field of study is usually referred to as 'coronal seismology' or 'solar magneto-seismology'. Here the observations do not cover the cyclotron-resonant regime, but are instead limited to the low-frequency (with periods of the order of minutes or longer) MHD regime. And the observations are not limited to the Alfvénic modes. In fact, since the solar atmosphere is highly structured, there are 'surface' and 'body' modes which differ from the usual MHD modes in a uniform medium; see especially Edwin & Roberts (1982, 1983) and Roberts (2000). For recent reviews of coronal seismology see Ballai & Forgács-Dajka (2010), De Moortel & Nakariakov (2012), Nakariakov & Verwichte (2005), Ofman (2009), Ruderman & Erdélyi (2009) and Wang (2011). And see Arregui et al. (2013), Asensio & Arregui (2013), Ballai & Orza (2012), Goossens et al. (2012), Jain

& Hindman (2012), Jelînek & Karlicky (2009), Li et al. (2013), Luna-Cardozo et al. (2012), Morton et al. (2011), Pascoe et al. (2013), Scott & Ruderman (2012) and Verwichte et al. (2013a,b) for some more recent papers.

In this paper we will consider one of the modes which has been particularly useful in coronal seismology, viz. the axisymmetric slow body mode on a thin cylindrical tube of magnetized plasma (see the aforementioned review papers as well as Bo et al. 2013; Luna-Cardozo et al. 2012; Macnamara & Roberts 2010, 2011; Telloni et al. 2013; Wang 2011); this mode is often referred to as the 'sausage mode', in which axisymmetric expansions and contractions of the magnetic field lines, and compressions and rarefactions of the plasma, propagate along the tube. If the tube is thin, the oscillations hardly disturb the surroundings, and thus the perturbation of total (i.e. magnetic plus thermal) pressure must be nearly zero. In the limit of zero tube radius the total pressure perturbation vanishes, and the sausage mode then propagates according to equation (17) below, with the relevant quantities evaluated inside the thin tube (Hollweg 1990). Equation (17) is also the dispersion relation for a planar MHD slow mode propagating highly obliquely to the background magnetic field; it is easy to show that the total pressure perturbation vanishes also for the plane wave in this limit. Our analysis will assume planar waves, but in the limit of highly oblique propagation it will also be valid for the sausage mode on a thin coronal structure. And of course our analysis will then apply also to highly oblique slow modes throughout interplanetary space, and not just on thin coronal structures. We note that there is a strong tendency for both waves and turbulence in the corona and solar wind to be characterized by high ratios of perpendicular to parallel wavenumbers, so our results can apply to a wide range of disturbances in space.

The new feature here will be the inclusion of one ion species drifting relative to the protons, in particular $\alpha$-particles, which are the most abundant 'heavy ion' in the solar atmosphere and solar wind. Allowing for drift is essential, since $\alpha$-particle drift is the rule rather than the exception in the regions of the solar wind which have been directly sampled by spacecraft (e.g., Marsch 1982; Reisenfeld et al. 2001; Kasper et al. 2013). Moreover, optical observations show that other heavy ions flow faster than the protons close to the Sun in the regions where the solar wind becomes supersonic (e.g. Cranmer 2009; Kohl et al. 1998), and there is no reason to expect that $\alpha$-particles will be any different; in fact some theoretical models (Isenberg 1990) of coronal heating and solar wind acceleration predict rapid $\alpha$-particle drifts close to the Sun.

We expect that our results will be relevant for interpreting remote observations of the slow mode, especially the sausage mode, close to the Sun. We will find that even non-drifting α-particles affect the dispersion relation in a non-simple way. The compressions and velocity fluctuations of the α-particles (or other heavy ions which might be observed remotely) can even have opposite signs compared to the protons, a fact that can prove useful in wave diagnostics. Our analysis will show that drifting α-particles can even drive the slow mode unstable; since the derivation uses the fluid equations the instability is non-resonant. But damping due to collisions (see Macnamara & Roberts 2010) and kinetic effects such as Landau and transit-time damping (Barnes 1966, 1967, 1968, 1969a,b), will probably eliminate the instability for typical conditions in the corona and solar wind; we present a fully kinetic calculation in Section 4.

The slow mode has not received a great deal of attention in studies of the solar wind, but we suggest that its importance may be underestimated. The instability mentioned in the previous paragraph could limit the α-particle drift in the solar wind, and thereby effect a conversion of the α-particle drift energy into thermal energy. In a low-beta plasma the drift threshold can be lower than the threshold for electromagnetic instabilities (Verscharen and Chandran 2013). Moreover, the slow mode may be involved in MHD turbulence, which we have already mentioned is a leading candidate for dissipating wave energy and heating the corona and solar wind (Chandran 2003, 2005; Cranmer & van Ballegooijen 2012; Kumar et al. 2006; Matsumoto & Suzuki 2012; Schekochihin et al. 2009; Verscharen et al. 2012); there is in fact some observational evidence for the presence of the slow mode in MHD turbulence (Howes et al. 2012; Klein et al. 2012; Smith & Zhou 2007, Yao et al. 2011). Slow mode waves may also be produced in the solar wind via parametric instability of the 'background' Alfvén waves, and the parametric instabilities may themselves play a role in the development of turbulence (Araneda et al. 2009; Del Zanna et al. 2001; Del Zanna & Velli, 2002; Ghosh et al. 1993; Goldstein 1978; Gomberoff 2006a,b,c, 2007a,b, 2008, 2009; Gomberoff et al. 1996, 2002, 2010; Hollweg et al. 1993; Hollweg 1994; Inhester 1990; Jayanti & Hollweg 1993a,b, 1994; Malara & Velli 1996; Vasquez 1995; Viñas & Goldstein 1991). Moreover, our analysis of the behavior of ions in the highly oblique slow wave might be used to distinguish true non-propagating pressure-balanced structures (PBSs) (Bruno & Carbone 2013; Tu & Marsch 1995; and references therein) in the solar wind from the propagating waves analyzed here, even though both exhibit constant total pressure. (A note on nomenclature: PBSs can be regarded as arising from the slow mode in the $k_z \rightarrow 0$ limit, while in our calculation the highly-oblique limit takes $k_y \rightarrow \infty$ while keeping $k_z$ finite and the waves propagate.) Finally, our analysis shows that the slow mode can produce temporal changes of the solar wind composition which are not of solar origin (c.f. Reisenfeld et al. 1999).

Before proceeding we offer one caveat. The highly oblique slow mode has vanishing total pressure perturbations, and this has been used to identify its presence in the solar wind (Howes et al. 2012; Smith & Zhou 2007). Vanishing total pressure pertubations are also associated with non-propagating PBSs. However, a fluid analysis suggests that this can also be a property of kinetic Alfvén waves; see equation (14) of Hollweg (1999). It is thus possible that associating vanishing total pressure perturbations just with the slow mode or PBSs can be erroneous. It is our expectation that the present analysis of the behavior of ions in slow waves might lead to an observational resolution of the ambiguities.

The analysis is found in Section 2. A multi-fluid approach is used, but we neglect collisions between plasma species as well as viscosity and heat conduction. Kinetic effects associated with the Landau resonance are also omitted in the multi-fluid study. We use expansions in (gyro frequencies)[-1] to restrict the analysis to the MHD regime, and we take the electrons to be massless but with finite pressure. We ignore the Alfvén mode but retain both the fast and slow modes; their dispersion relation is further restricted to the slow mode by going to the highly oblique limit (equations 15a and 18). Section 3 presents some numerical examples, showing in particular the instability driven by $\alpha$-drift; some analytical results for the instability are presented in equations (26, 29 and 30). Section 4 discusses kinetic effects and associates the instability with its kinetic counterpart. Section 5 summarizes our results.

## 2. Analysis

We consider a plane wave propagating in a uniform background plasma consisting of electrons, protons, and one ion species immersed in a uniform magnetic field $\mathbf{B_0} = (0,0,B_0)$ pointing in the $z$-direction (the subscript '0' will denote a background quantity). Any background particle flows $\mathbf{V_0} = (0,0,V_0)$ are uniform and along $\mathbf{B_0}$. All fluctuations (denoted by prefix 'δ') are assumed to be small and to vary as $\exp[i\,k_y\,y + i\,k_z\,z - i\,\omega\,t]$. We consider low-frequency fluctuations in the MHD regime. The wave magnetic field is taken to lie in the $y$-$z$ plane, i.e. $\delta B_x = 0$; this eliminates the MHD Alfvén mode but retains the fast and slow modes. Consistent with this choice and with MHD we take $\delta E_y = 0$ (**E** denotes the electric field) and there are no $\delta \mathbf{E} \times \mathbf{B}_0$ drifts in the $x$-direction. However, in order for there to be charge quasineutrality we still need to allow for a (small) $\delta E_z$. (In Section 4 we will revisit the assumptions $\delta B_x = 0$ and $\delta E_y = 0$.)

After linearization, the momentum equation for each species (charge $q$, mass $m$, and background concentration $n_0$) is

$$-i\,\omega'\,\delta \mathbf{V} = \frac{q}{m}\left( \delta \mathbf{E}' - \frac{i\mathbf{k}\,\delta p}{q\,n_0} + \frac{\delta \mathbf{V} \times \mathbf{B}_0}{c} \right) \qquad (1)$$

where $c$ is the speed of light (cgs units will be used throughout), $\omega' = \omega - k_z\,V_0$, $p$ is plasma pressure (for adiabatic fluctuations $\delta p = m\,v_s^2\,\delta n$ with $v_s$ the sound speed for each species) and $\delta \mathbf{E}' = \delta \mathbf{E} + \mathbf{V}_0 \times \delta \mathbf{B}/c$. With the assumed polarization of $\delta \mathbf{B}$, the Lorentz term affects only $\delta E_x'$:

$$\delta E_x' = (\omega'/\omega)\,\delta E_x \qquad (2)$$

where use has been made of Faraday's law.

After linearization, the mass conservation equation for each species is

$$\omega'\,\delta n = n_0\left( k_y \delta V_y + k_z \delta V_z \right) \qquad (3)$$

Equation (1) is first solved for the three components of $\delta \mathbf{V}$ in terms of $\delta \mathbf{E}'$ and $\delta n$. The components $\delta V_y$ and $\delta V_z$ are then inserted into equation (3) which can then be solved for $\delta n$ for each species:

$$\delta n = \frac{i\,q\,n_0\left[ i\,k_y\,\omega'\,\Omega\,\delta E_x' + k_z\,\delta E_z(\Omega^2 - \omega'^2) \right]}{m\left[ \Omega^2(\omega'^2 - k_z^2 v_s^2) - \omega'^2(\omega'^2 - k^2 v_s^2) \right]} \qquad (4)$$

where $\Omega = q\,B_0\,/(m\,c)$ is the cyclotron frequency. Equation (4) is then used to eliminate $\delta n$ from the prior result for $\delta \mathbf{V}$. The resulting expressions (which will not be written down) for the three components of $\delta \mathbf{V}$ are then expanded in terms of $1/\Omega$ and we obtain

$$\delta V_x\,/\,c = \frac{k_y\,k_z\,v_s^2\,\delta E_z - i\,\omega'(\omega'^2 - k^2 v_s^2)\delta E_x'\,/\,\Omega}{B_0(\omega'^2 - k_z^2 v_s^2)} \qquad (5)$$

$$\delta V_y\,/\,c = -\delta E_x'\,/\,B_0 \qquad (6)$$

$$\delta V_z\,/\,c = \frac{i\,\omega'\,\Omega\,\delta E_z - k_y\,k_z\,v_s^2\,\delta E_x'}{B_0(\omega'^2 - k_z^2 v_s^2)} \qquad (7)$$

where $k^2 = k_y^2 + k_z^2$. In equations (5)-(7) we have kept the term proportional to $1/\Omega$ only in $\delta V_x$, since that term gives the polarization drift which we will only need when we calculate the fluctuation current in the x-direction, $\delta j_x$. And for consistency we expand equation (4) in terms of $1/\Omega$; the two leading terms are

$$\frac{\delta n}{c\,n_0} = \frac{i\,k_z\,\Omega\,\delta E_z - k_y\,\omega'\,\delta E_x'}{B_0(\omega'^2 - k_z^2 v_s^2)} \qquad (8)$$

For protons and ions, the quantity in parentheses in the denominators of equations (7-8) does not constitute a singularity, since it cancels out when $\delta E_z$ is replaced with equations (10-11) below. However, numerical examples reveal that there are a few points in solution space where the denominators of equations (5, 7 and 8) pass through zero. Dropping the second term in the denominator of equation (4) is then not valid, but since $\Omega$ is large this is a problem only for extremely small ranges of $V_0$ surrounding those points. Including the dropped term would take us out of the realm of MHD, and we simply note the existence of this difficulty. We note too that equation (6) omits a term proportional to $\delta E_z\,/\,[\Omega(\omega'^2 - k_z^2\,v_s^2)]$. Even though $\delta E_z$ is generally small, this is invalid when the denominator passes through zero, or at certain points in solution space where $\delta E_z \rightarrow \infty$. Again, including the dropped term would take us out of the realm of MHD. And including the dropped term would not affect the dispersion relation, which will be derived without use of equation (6).

For electrons we assume $m_e \rightarrow 0$, $\left| \Omega_e \right| \rightarrow \infty$, $v_{se}^2 \rightarrow \infty$, keeping in mind that $v_{se}^2 / \Omega_e$ remains finite. Subscripts $e$, $p$ and $i$ will denote electrons, protons, and ions (sometimes including the protons), respectively. Thus

$$\delta n_e = \frac{i \, e \, n_{0e} \, \delta E_z}{\gamma \, K \, T_e \, k_z} \tag{9}$$

where $e$ is the proton charge, $\gamma$ is the usual ratio of specific heats, $K$ is Boltzmann's constant, and $T$ is temperature; equation (9) also follows directly from the z-component of the electron momentum equation with $m_e = 0$. (To ease the mathematics we take $\gamma$ to be the same for all species. But since $\gamma$ is always multiplied by a temperature, at the end we can simply replace $\gamma T_{e,p,i}$ with $\gamma_{e,p,i} T_{e,p,i}$.)

We now use equation (8) and plasma quasineutrality, viz. $e \, \delta n_e = \sum q_i \, \delta n_i$, where the summation is over protons and ions, to obtain $\delta E_z$:

$$\delta E_z = \frac{i \, c \, k_y \, \delta E_x}{B_0 \, D \, \omega} \sum \frac{q_i \, n_{0i} \, \omega_i'^2}{\omega_i'^2 - k_z^2 v_{si}^2} \tag{10}$$

where

$$D = \frac{e^2 n_{0e}}{\gamma \, K \, T_e \, k_z} - k_z \sum \frac{q_i^2 n_{0i} / m_i}{\omega_i'^2 - k_z^2 v_{si}^2} \tag{11}$$

and we have used equation (2). Equations (10-11) show that $\delta E_z$ is of the order of $k_y \, \omega \, \delta E_x / (k_z \, \Omega_{i,p})$ which is consistent with the ordering of terms in equations (5-8).

Having $\delta E_z$ in terms of $\delta E_x$ we can now compute the particle velocities in terms of $\delta E_x$. We are particularly interested in $\delta V_x$, equation (5), in order to compute the electric current $\delta j_x$. For the electrons with $m_e \rightarrow 0$ we have

$$\delta V_{xe} = -\frac{c\,k_y\,\delta E_z}{k_z\,B_0} \qquad (12)$$

With the help of equation (9) it is easy to show that this is simply the drift arising from $\nabla \delta p_e$ in equation (1). For protons and ions we use equations (5) and (2). With the help of Ampere's and Faraday's laws, and the supplemental condition $k_y \delta B_y + k_z \delta B_z = 0$, we have

$$\delta j_x = -e\,n_{0e}\,\delta V_{xe} + \sum q_i\,n_{0i}\,\delta V_{xi} = -\frac{i\,k^2 c^2 \delta E_x}{4\,\pi\,\omega} \qquad (13)$$

where the sum is again over protons and ions. When $\delta E_x$ is eliminated, the second equality gives the dispersion relation.

As an illustration and a check of our procedure, consider a plasma consisting of massless electrons and one species of drifting ions; charge neutrality requires $e\,n_{0e} = q_i\,n_{0i}$. Following the recipe outlined above, we obtain the dispersion relation

$\omega'^2(\omega'^2 - k^2\,v_{s,tot}^2) - k^2 v_A^2(\omega'^2 - k_z^2\,v_{s,tot}^2) = 0$, where $v_A$ is the Alfvén speed based on the mass density of the drifting ions and $v_{s,tot}^2 = [\gamma\,K\,T_i + (n_{0e}\,/\,n_{0i})\gamma\,K\,T_e]\,/\,m_i$. This is immediately recognized as the usual dispersion relation for the fast and slow modes convected by the drifting ions.

Adding non-drifting protons to the above example (thus we are working in the proton frame), but still with the constraint of quasineutrality $e\,n_{0e} = e\,n_{0p} + q_i\,n_{0i}$, gives the dispersion relation:

$$k^2 C_k + k_z^2 C_{kz} = 0 \qquad (14)$$

where

$$C_k = (1 + \beta_{ep} + AN\beta_{ei})(1 + NZ)v^2v'^2 - [(1 + \beta_{ep})\beta_i + NZ(1 + \beta_p)\beta_{ei}]v^2 \tag{15a}$$

$$-[(1 + AN\beta_i)\beta_{ep} + NZ(1 + AN\beta_{ei})\beta_p]v'^2 + \beta_{ep}\beta_i + NZ\beta_{ei}\beta_p$$

$$C_{kz} = -(1 + NZ)v^4v'^2 - AN(1 + NZ)v^2v'^4 + (\beta_i + NZ\beta_{ei})v^4 \tag{15b}$$

$$+ AN(\beta_{ep} + NZ\beta_p)v'^4 + 2AN(\beta_i - \beta_{ei})v^2v'^2$$

and $v = \omega/(k_z v_{Ap})$, $v' = \omega'/(k_z v_{Ap}) = v - v_0$, $v_0 = V_0/v_{Ap}$, $\beta_{p,i} = \gamma K T_{p,i}/(m_{p,i}v_{Ap}^2)$,

$\beta_{ep} = \gamma K(T_e + T_p)/(m_p v_{Ap}^2)$, $\beta_{ei} = \gamma K(T_i + ZT_e)/(m_i v_{Ap}^2)$, $N = n_{0i}/n_{0p}$, $A = m_i/m_p$,

$Z = q_i/e$, and $v_{Ap}$ is the Alfvén speed based on the mass density of the protons. With three

temperatures and four $\beta$'s, only three of the $\beta$'s are independent; for example

$$\beta_{ei} = (Z/A)(\beta_{ep} - \beta_p) + \beta_i \tag{16}$$

For the reasons outlined in the Introduction we are going to limit our attention to the highly oblique slow mode. This mode has $\omega$ and $\omega'$ proportional to $k_z$ even as $k \to \infty$; for example, for highly oblique propagation the familiar MHD slow mode obeys

$$\omega^2/k_z^2 \approx v_A^2 v_s^2/(v_A^2 + v_s^2) \tag{17}$$

We recognize approximation (17) as giving the dispersion relation for the sausage mode on a thin cylindrical structure (Edwin and Roberts 1983; Hollweg 1990; Roberts 2000); the square root of the right-hand side is sometimes called 'the cusp speed'. From equation (14) and the definitions of $C_k$ and $C_{kz}$ we see that this can only happen if

$$C_k = 0 \tag{18}$$

and this is the desired dispersion relation. The advantage here is that the quantity $C_k$ is a quartic in $v$ and thus equation (18) is analytically solvable using standard packages.

For $N \to 0$, and $v_0 = 0$, the leading terms for the four roots of (18) are

$$v^2 \approx \beta_{ep}/(1 + \beta_{ep}) \equiv v^2_{\beta ep} \tag{19a}$$

and

$$v^2 \approx \beta_i \equiv v^2_{\beta i} \tag{19b}$$

(If $N = 0$, the second set of roots is extraneous.) Approximation (19a) is equivalent to approximation (17) for an electron-proton plasma. For $v_0 = 0$ and $v_{\beta i} \ll v_{\beta ep}$, a small-$N$ expansion of equation (18) gives the following correction to equation (19a):

$$v^2 \approx v^2_{\beta ep} - \frac{N Z v^4_{\beta ep}(\beta_{ep} - \beta_p)}{A \beta^3_{ep}} \Big[ A\beta_{ep}(1 + 2\beta_{ep}) - Z(1 + \beta_{ep})(\beta_{ep} - \beta_p) \Big] \tag{20}$$

With $A > Z$ the correction is always negative; it can be shown that increasing $\beta_i$ reduces the magnitude of the correction. On the other hand, if $v_{\beta i} \gg v_{\beta ep}$ the small-$N$ correction to equation (19a) is

$$v^2 \approx v^2_{\beta ep} - N v^6_{\beta ep} \Big[ A + Z(1 + \beta_{ep})(\beta_{ep} - \beta_p)\beta^{-3}_{ep} \Big] \tag{21}$$

and the correction is always negative. (Equation (16) has been used in the previous two equations.) For $\alpha$-particles and the moderate values of $\beta_p$ and $\beta_{ep}$ to be used in Figure 1, the corrections are 8% at low $\beta_i$ and 11% at high $\beta_i$, while for the low values of $\beta_p$ and $\beta_{ep}$ to be used in Figure 4 the corrections are only 2.5-3%. If $v_{\beta i} \approx v_{\beta ep}$ the expansion becomes singular because the roots in equations (18) are not distinctly separated, but full numerical solutions show that the corrections do become larger as $v_{\beta i} \rightarrow v_{\beta ep}$ from either side. We will henceforth refer to modes which are close to equations (17) and (19a) as 'cusp-like' and we will refer to solutions which are close to equation (19b) as 'ion-dominated'. The phase speed of cusp-like solutions approximately corresponds to the ion-acoustic speed, and the phase speed of ion-dominated solutions corresponds to the ion-sound speed (Mann et al. 1997). The modes can propagate in both directions with respect to the background magnetic field, and while the phase speed of the ion-acoustic mode is barely affected by the drifting ions, the ion-sound wave propagates with the ion-sound speed in the reference frame in which the ions are at rest (Verscharen & Marsch 2011).

As a further check, we have also derived dispersion relation (18) by starting with the fact that for the MHD slow mode the perturbation of the total pressure, $\delta p_{tot}$, is zero in the limit of highly oblique propagation. As a concrete illustration of this assertion, for the standard MHD slow mode in a single fluid with sound speed $v_s$ it is easy to derive

$$\frac{\delta p}{\delta p_{mag}} \approx -1 - \frac{k_z^2}{k_y^2 (1+\beta)} \qquad (22)$$

with $\beta = v_s^2 / v_A^2$ and $\delta p_{mag}$ is the magnetic pressure perturbation. Thus $\delta p_{tot} \rightarrow 0$ asymptotically as $k_y / k_z \rightarrow \infty$. The derivation of the dispersion relation in this manner is straightforward since equations (8-10) can be used to derive $\delta p$ in terms of $\delta E_x$, and Faraday's law can be used to write the magnetic pressure perturbation in terms of $\delta E_x$. Thus $\delta E_x$ can be eliminated and the dispersion relation (18) results.

One of the interesting, and perhaps surprising, features of the highly oblique slow mode is that $\delta n_i$ and $\delta n_p$ can be $\pi$ radians out of phase; we will show in Figure 3 that this can be true even when there is no ion drift. (This was in fact the original motivation for this study. Viall et al. (2009) reported anti-correlated proton and $\alpha$-particle variations in the solar wind. We thought these observations might be due to waves, but in retrospect a coronal origin seems more likely. However, the principle still holds: the slow mode can temporally and locally change the plasma composition.) Using equations (8), (10) and (11) we obtain

$$\frac{\delta n_i / n_{0i}}{\delta n_p / n_{0p}} = \frac{(\beta_{ei} - \beta_i)v^2 - (v - v_0)^2 \left[\beta_{ep} + NZ\beta_p - v^2(1 + NZ)\right]}{-(\beta_i + NZ\beta_{ei})v^2 + (v - v_0)^2 \left[AN(\beta_{ei} - \beta_i) + v^2(1 + NZ)\right]} \quad (23)$$

We will show in Figure 3 that $\delta n_i$ and $\delta n_p$ can have opposite signs, even when there is no ion drift.

Another usesful diagnostic of the wave properties is the ratio $\delta V_{zi} / \delta V_{zp}$, which can be derived using equations (7) and (10)-(11):

$$\frac{\delta V_{zi}}{\delta V_{zp}} = \frac{v - v_0}{v} \frac{\{A\beta_i(\beta_i - \beta_{ei}) + Z[\beta_{ei}(NZv^2 + v^2 - NZ\beta_p) - \beta_i\beta_p]\}}{\{A(\beta_i - \beta_{ei})[\beta_i - (1 + NZ)(v - v_0)^2] + Z\beta_p[-\beta_i - NZ\beta_{ei} + (1 + NZ)(v - v_0)^2]\}} \quad (24)$$

We will show in Figure 2 that $\delta V_{zi}$ and $\delta V_{zp}$ can have opposite signs, even when there is no ion drift. Since the ions and protons do not move together, the phase speed cannot be calculated from equation (17) using a value for $v_A$ based on the total mass density and a value for $v_s$ using the full plasma pressure and the total mass density. (This can be easily verified with some numerical examples.)

## 3. Numerical Examples

Our numerical examples will all consider a plasma consisting of electrons, protons, and α-particles ($A = 4$, $Z = 2$). For our first example we take $T_e = 10^5 K$, $T_p = 8 \times 10^4 K$, $T_i = 3T_p$, $v_{Ap} = 40$ km s$^{-1}$, and $N = 0.06$. Although mainly illustrative, these parameters correspond roughly to conditions in the solar wind at 1 AU. We also somewhat arbitrarily take $\gamma = 1.3$ rather than 5/3, in the expectation that this may represent internal heating processes including electron heat conduction. In this case $v\beta_{ep} = 0.74$ and $v\beta_i = 0.635$. The $\beta$'s are not small; for example $\beta_{ep} = 1.21$. Figure 1 plots the real parts of the four roots of $v$ obtained from equation (18) vs. $v_0$. (In Figures 1 and 4, points where the denominators of equations (5, 7 and 8) pass through zero are indicated by '×' and an arrow for protons and ions respectively.) The roots which are almost horizontal lines are cusp-like and close to $v \approx \pm v_{\beta ep}$; the roots which are almost diagonal lines are ion-dominated and are close to

$$v \approx v_0 \pm v_{\beta i}. \qquad (25)$$

The gray line, $v = v_0$, shows that the ion-dominated modes are indeed largely convected by the drifting ions. There is an avoided crossing around $v_0 \approx 0.1$ where the roots can not be simply characterized. (We will see a similar avoided crossing in Figure 4. Both Figures 1 and 4 have $v_{\beta ep} > v_{\beta i}$. We will not show an example in which the inequality is reversed, but in such cases the avoided crossing occurs for the modes for which $v < 0$.) For $1.3 < v_0 < 1.4$ there are two real roots and two complex conjugate roots, one of which is unstable; the dimensionless growth rate (multiplied by 10) is shown as the thin curve. The instability occurs when the forward-propagating cusp-like wave has about the same phase speed as the backward-propagating (in the ion frame) ion-dominated mode. When the phase speeds are about equal, both modes couple, and the solutions collapse to two complex solutions with equal real parts of the wave frequency. Because the $\beta$'s are not small, the instability requires $V_0 > v_{Ap}$, and other electromagnetic instabilities with larger growth rates (Verscharen & Chandran 2013) will probably be more important. However, using the small-$N$ approximations for the roots, the root intersection leading to instability will occur close to

$$v_0 = v_{\beta ep} + v_{\beta i} \qquad (26)$$

so that in a plasma with low $\beta$'s instability can occur for $V_0 << v_{Ap}$ where electromagnetic modes should be stable; we will show such an example in Figure 4.

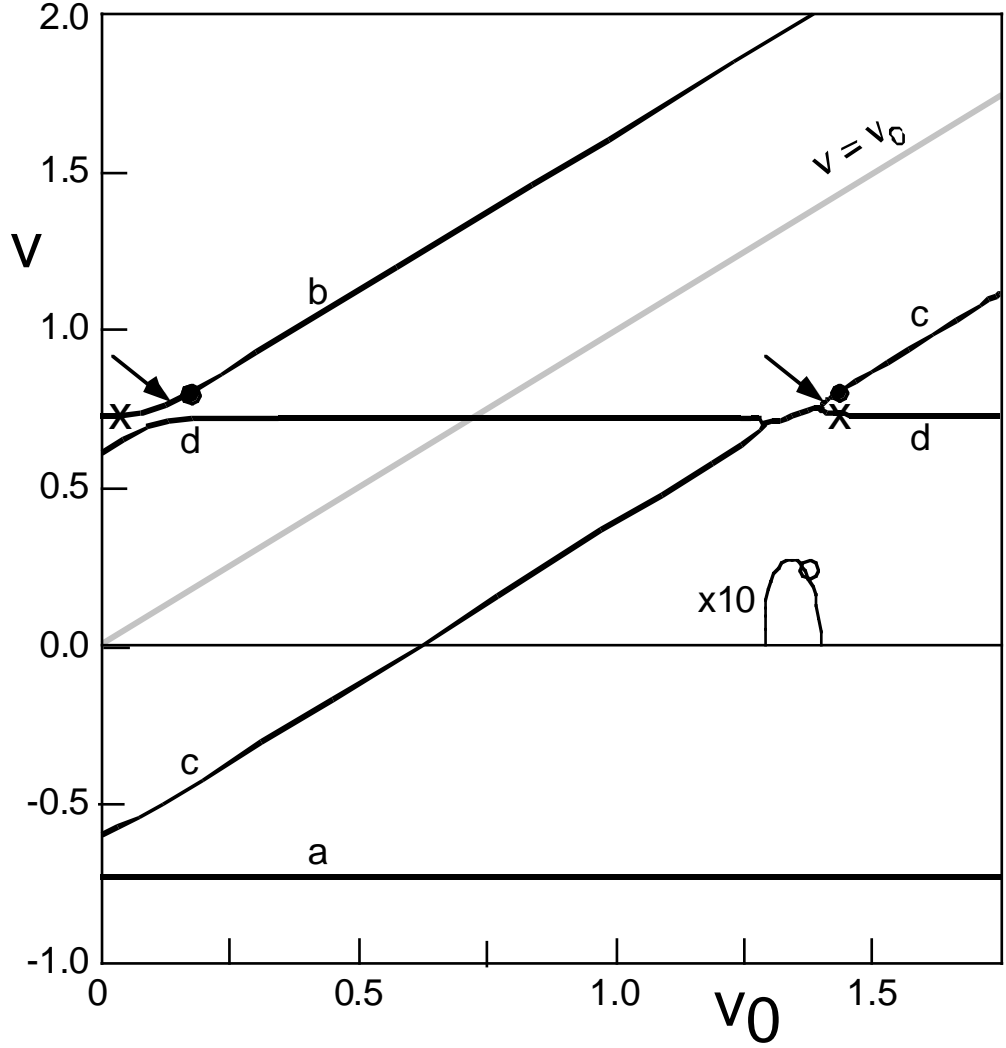

**Figure 1**. The heavy black curves give the real parts of the four roots (designated a-d) of equation (18) vs. normalized ion drift speed $v_0$. Nearly horizontal curves are the cusp-like modes while diagonal curves are the ion-dominated modes; the gray line, $v = v_0$, shows that the ion-dominated modes are largely convected by the drifting ions. The thin curve gives the growth rate multiplied by 10; the corresponding damping rate is not shown. The chosen parameters $T_e = 10^5 K$, $T_p = 8 \times 10^4 K$, $T_i = 3T_p$, $\gamma = 1.3$, $v_{Ap} = 40$ km s$^{-1}$ and $N = 0.06$ are roughly representative of the solar wind at 1 AU. Points where $\omega'^2 - k_z^2 v_s^2 = 0$ are indicated by '$\times$' and arrows for protons and ions respectively. The solid circles denote solution points where $D = 0$ as given by equations (27-28). The open circle is the approximate maximum growth rate as given by equation (29).

Some insight into the nature of the modes can be gained by examining $\delta V_{zi} / \delta V_{zp}$. This ratio spans a large range of values, so in order to fit the essential information into one graph we plot the quantity $Sgn[\text{Re}(\delta V_{zi} / \delta V_{zp})]\ Abs[\delta V_{zi} / \delta V_{zp}]^{1/4}$, which is shown in Figure 2 for the same parameters used in Figure 1. Look first at curves 'a' and 'b' in the lower-left part of the figure. Comparing with Figure 1 (which uses the same labelling), as $v_0 \to 0$ they represent $\delta V_{zi} / \delta V_{zp}$ for the cusp-like modes. But note that not only do the ions and protons move at different speeds, but they are $\pi$ radians out of phase. This supports our earlier statement that equation (17) is inaccurate even when $v_0 = 0$. Look next at curves 'a' and 'd' which correspond to the cusp-like modes when there is significant drift, $v_0 > 0.2$ or so. We see that the motions are dominated by the protons (as expected) with $|\delta V_{zi} / \delta V_{zp}| < 1$. Note too that in Figure 1 root 'd' has $v_0 = v$ at $v_0 \approx 0.72$; the ions are then moving with the wave, and $\delta V_{zi} = 0$. Finally, consider the solutions given by curve 'c', most of curve 'b', and the left portion of curve 'd' in Figure 1. We have already claimed that they are dominated by the ion dynamics; Figure 2 confirms this by showing $|\delta V_{zi} / \delta V_{zp}| > 1$ for these roots. Note too that the curve 'c' in Figure 1 has $v = 0$ when $v_0 \approx 0.61$; this leads to the infinity in Figure 2 for reasons which will be discussed in connection with Figure 3. Finally, the two complex conjugate roots, which appear when $1.3 < v_0 < 1.4$, have nearly constant modulus but varying phase in this range.

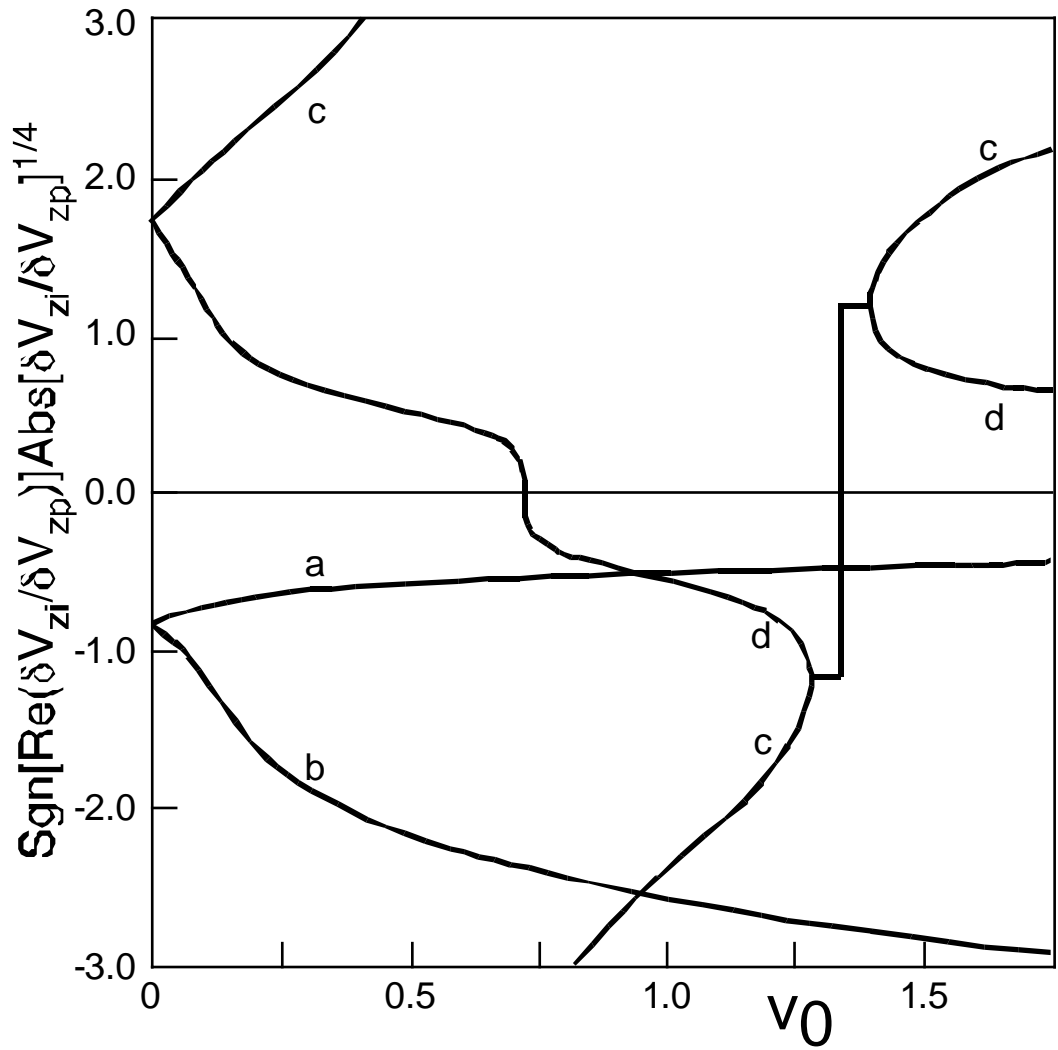

**Figure 2.** Scaled representation of $\delta V_{zi} / \delta V_{zp}$ for the four roots in Figure 1. Since protons and ions do not move together equation (17) is inaccurate even when $v_0 = 0$.

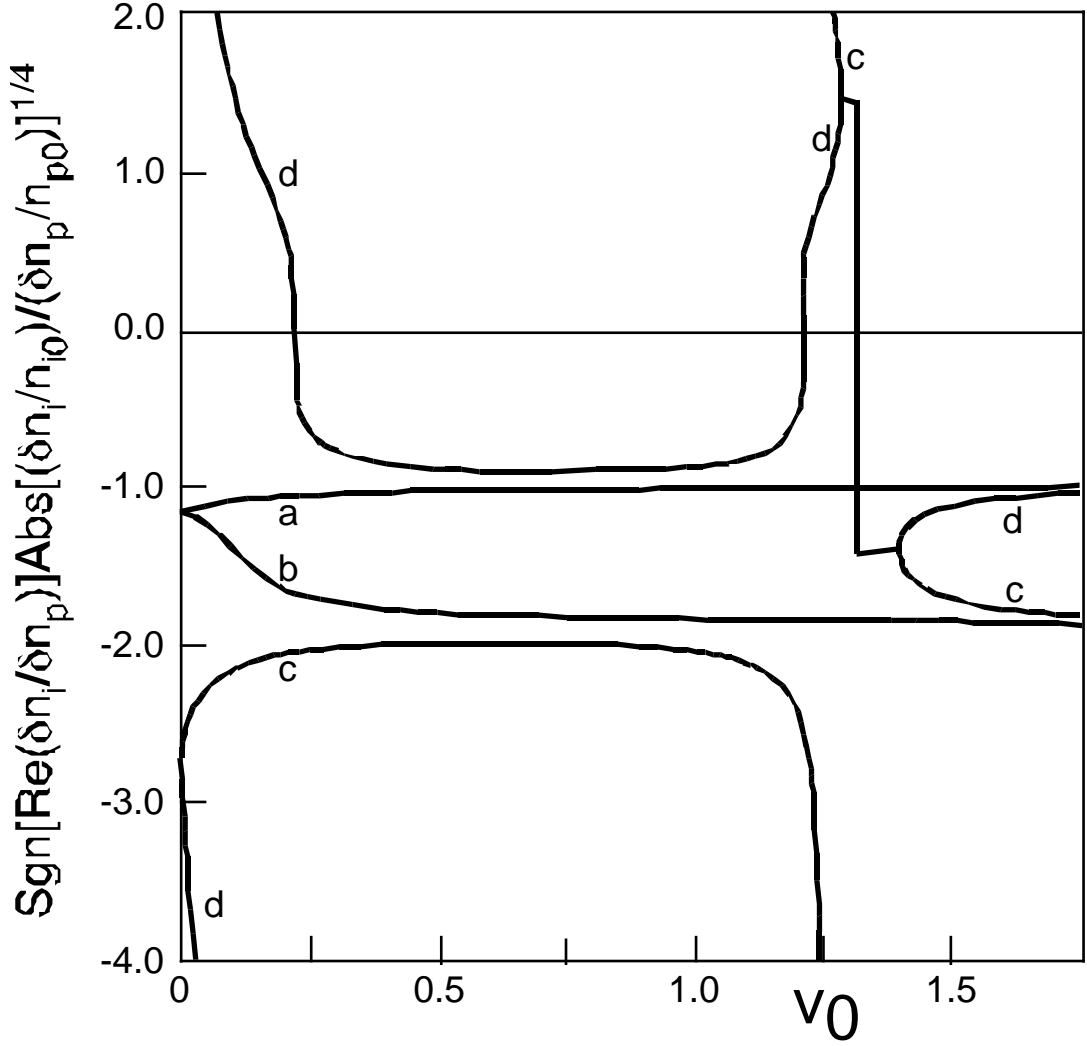

**Figure 3**. Scaled representation of $(\delta n_i / n_{i0})/(\delta n_p / n_{p0})$ for the four roots in Figure 1.

Figure 3 displays the density ratio $(\delta n_i / n_{i0})/(\delta n_p / n_{p0})$ for the same case. (As in Figure 2 we plot the sign of the real part multiplied by the absolute value raised to the ¼ power.) There is little resemblance to Figure 2 for $\delta V_{zi} / \delta V_{zp}$; the reason is that $\delta n$ is influenced also by $(k_y \, \delta V_y)$, according to equation (3). Roots 'c' and 'd' give infinities where $\delta n_p \rightarrow 0$, while root 'd' gives two zeroes where $\delta n_i = 0$. Note that even for $v_0 = 0$ the two cusp-like roots 'a' and 'b' have proton and ion density fluctuations which are opposite in sign. Finally, the cusp-like roots have modest values of $(\delta n_i / n_{i0})/(\delta n_p / n_{p0})$, while the ion-dominated roots have larger values of $(\delta n_i / n_{i0})/(\delta n_p / n_{p0})$.

The two filled circles in Figure 1 indicate where $D \to 0$ and $|\delta E_z| \to \infty$; consequently $\delta V_{zi}$, $\delta V_{zp}$, $\delta n_i$ and $\delta n_p$ all become singular there, but the ratios shown in Figures 2 and 3 stay finite. According to equation (10), $\delta E_z$ also has a singularity where $\omega = 0$. This occurs where the ion-dominated root 'c' crosses the horizontal axis in Figure 1, and $\delta V_{zi}$, $\delta n_i$ and $\delta n_p$ all become singular there, but $\delta V_{zp}$ stays finite in virtue of the $\omega'$ ($= \omega$ for protons) in the numerator of equation (7); this explains the singularity at $v_0 \approx 0.72$ in Figure 2.

After extensive algebra we have been able to show that the values of $v_0$ leading to $D = 0$ are given by

$$v_{0,D=0}^2 = \frac{(\beta_{ep}\beta_i + N Z \beta_{ei}\beta_p)\left[\beta_{ei} + \beta_{ep} \pm 2\sqrt{\beta_{ei}\beta_{ep}}\right]}{(1 + N Z)\beta_{ei}\beta_{ep}} \qquad (27)$$

The corresponding values of $v$ in Figure 1 are given by

$$v_{D=0}^2 = \frac{\beta_{ep}\beta_i + N Z \beta_{ei}\beta_p}{(1 + N Z)\beta_{ei}} \qquad (28)$$

(In Figures 1, and 4 below, the smaller value of $v_{0,D=0}$ is close to the avoided crossing, which in those Figures occurs at positive $v$, and thus the positive square root of Equation (28) must be used. But as mentioned in connection with Figure 1, when $v_{\beta i}$ is sufficiently large the avoided crossing occurs at negative $v$, and then the negative square root of (28) must be used.) Even though the two zeroes of $D$ occur at different values of $v_0$ in Figure 1, and on different branches of the dispersion relation, they have the same value of $|v|$; we are unable to offer a physical explanation of this result.

Figure 4 displays the solutions of the dispersion relation (18) for conditions roughly representative of the acceleration region of the fast solar wind. We take $T_e = 10^6 K$, $T_p = 3 \times 10^6 K$, $T_i = 4 T_p$, and $v_{Ap} = 1500$ km s$^{-1}$; the ion is again fully-ionized helium with $N = 0.06$. In this case the $\beta$'s are small (e.g. $\beta_{ep} = 0.02$), and the characteristic speeds are $v_{\beta ep} = 0.137$ and $v_{\beta i} = 0.12$. Figure 4 closely resembles Figure 1, except the two filled circles at which $D = 0$ are at qualitatively different locations. The main difference from Figure 1 is that the instability now occurs for $V_0 << v_{Ap}$, and there should be no competing electromagnetic instabilities. The growth rate (shown multiplied by 10) is again small, a little more than 3 percent of the real part of the frequency.

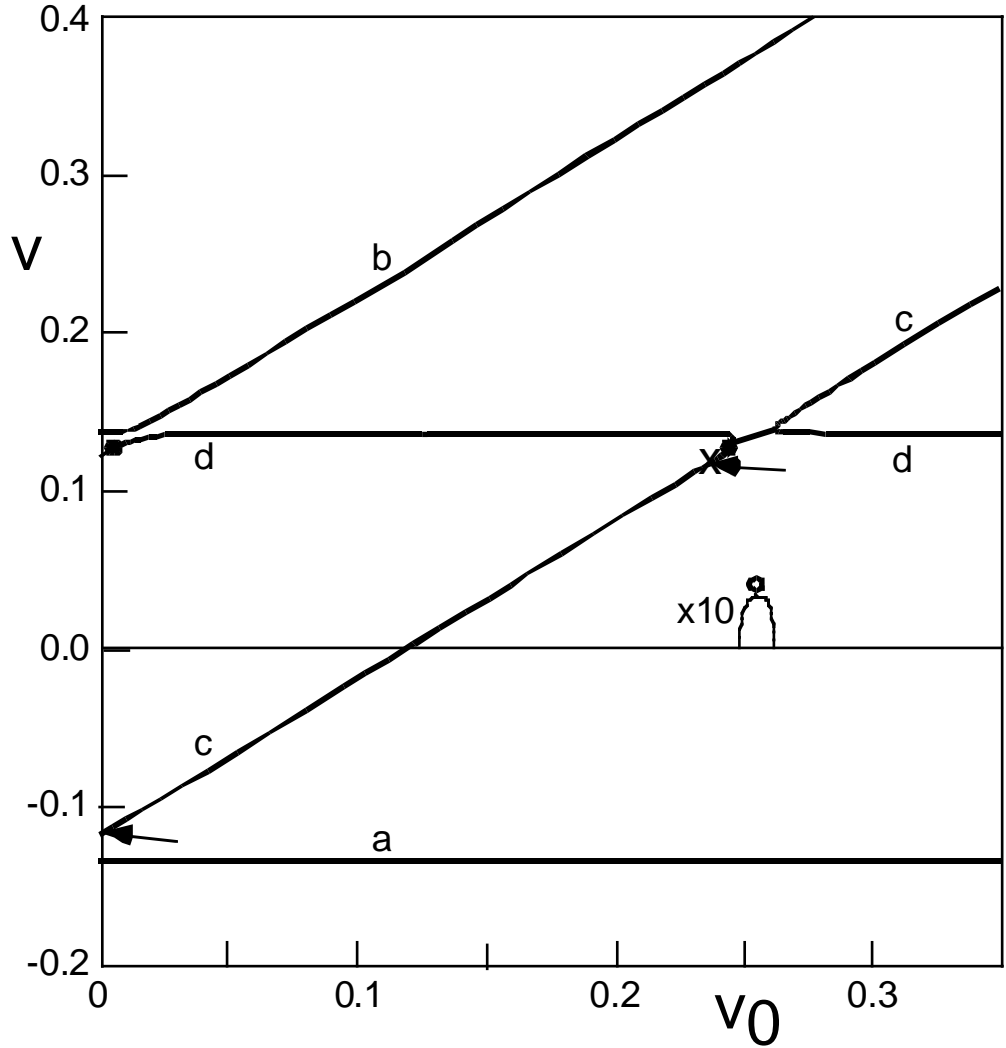

**Figure 4**. Same as Figure 1 but $T_e = 10^6 K$, $T_p = 3 \times 10^6 K$, $T_i = 4 T_p$, $\gamma = 1.3$, $v_{Ap} = 1500$ km s$^{-1}$ and $N = 0.06$.

We now seek a simple analytical approximation for the maximum growth rate of the instability which appears in Figures 1 and 4. We assume that the maximum growth rate occurs at the value of $v_0$ given by equation (26). We also assume that the real part of the normalized frequency $v$ in the unstable region is close to $v_{\beta ep}$ (approximation 19a), and we expand the dispersion relation (18) around that value of $v$, with $dv = v - v_{\beta ep}$. For the parameters considered in this paper we find that we can safely drop terms involving $dv^3$ and $dv^4$; it is somewhat less accurate but very convenient to drop the term proportional to $dv$. We also drop terms proportional to $N^2$. The maximum dimensionless growth rate can then be approximated as

$$v_{growth,max} \approx \frac{(AN)^{1/2} v_{\beta ep}^{3/2} \left| \beta_i - \left(1 - \beta_p / \beta_{ep}\right) Z / A \right|}{2 v_{\beta i}^{1/2}} \qquad (29)$$

(Equation (16) has again been used.) The point $(v_{\beta ep} + v_{\beta i}, 10\, v_{growth,max})$ is displayed as the open circles in Figures 1 and 4. Note that the growth rate can vanish when $\beta_i = (Z/A) T_e / (T_e + T_p)$, which is close to being fulfilled for the warm plasma case of Figure (1). But when $\beta_i$ substantially exceeds this value, the growth rate increases as $T_i^{3/4}$ and can become substantial even when $N$ is small; for example, for the parameters in Figure 1, but with $T_i$ increased to $8 T_p$, the maximum growth rate is 14% of the real part of $v$.

For small $\beta$'s and for modest values of $\gamma_p T_p / \gamma_e T_e$ (as in Figure 4), $\beta_i$ can be dropped from the absolute value in equation (29), and the maximum growth rate normalized to the real part of the frequency (approximated as $v_{\beta ep} \approx \beta_{ep}^{1/2}$) is then

$$\frac{v_{growth,max}}{v_{\beta ep}} \approx \left(\frac{N}{A}\right)^{1/2} Z \, \frac{\gamma_e T_e / (\gamma_p T_p)}{2[1 + \gamma_e T_e / (\gamma_p T_p)]^{3/4}} \left(\frac{\beta_p}{\beta_i}\right)^{1/4} \qquad (30)$$

In this case hot protons and hot ions, as seem to be found in coronal holes (e.g. Antonucci et al. 2000; Cranmer 2009; Kohl et al. 1998), reduce the normalized growth rate. For the parameters used in Figure 4 equation (30) gives a normalized growth rate of 0.033; for a wave with a period of 300 seconds, the e-folding time is then 1450 seconds, which is not unreasonably long.

It is not clear whether the ion-drift-driven instability would play a role in the corona or solar wind. Since the instability is driven by the ion drift, one would expect the instability to react back on the ions and limit their drift. In low-beta plasmas, the required ion drift can be substantially less that the drift needed to excite electromagnetic instabilities (Verscharen & Chandran 2013). But since the instability considered here exists over only a limited range of $v_0$, it is possible that external accelerating forces could push the ions through to higher stable values of $v_0$ before the instability can grow to large enough amplitudes to effectively react back on the ions. It appears that this problem might not occur for the cyclotron-resonant electromagnetic instabilities driven by ion drift. Moreover, if dissipation associated with the Landau resonance, or with viscosity or heat conduction which have not been included in our analysis, is large enough, the instability might be eliminated altogether. It is interesting to note, however, that the instability is Landau-resonant with the Landau-unstable part of the ion distribution function (this follows from equation (26) and the fact that $v_{\beta ep}$ is a good approximation for the normalized phase speed).

## 4. Kinetic Effects

The resonant kinetic counterpart of the described non-resonant fluid instability is the ion/ion-acoustic instability (Fried & Wong 1966; Gary & Omidi 1987; Gary 1993). The ion-acoustic mode is driven unstable by Landau-resonant beam ions. A Landau-resonant instability can only be driven if

$$\omega < k_z v_0 \qquad (31)$$

under the assumption that the beam ions have a Maxwellian distribution function (Verscharen & Chandran, 2013). Due to its polarization properties, the ion-acoustic mode is prone to strong Landau and transit-time damping by the plasma protons (and to a lesser extent the electrons). But since the phase speed of the ion-acoustic mode increases as $(T_e + T_p)^{1/2}$, the resonant damping can be reduced if the electron temperature is large enough so that only the few protons in the tail of their distribution can resonate with the wave at the same wavenumber at which the ions resonantly drive the instability.

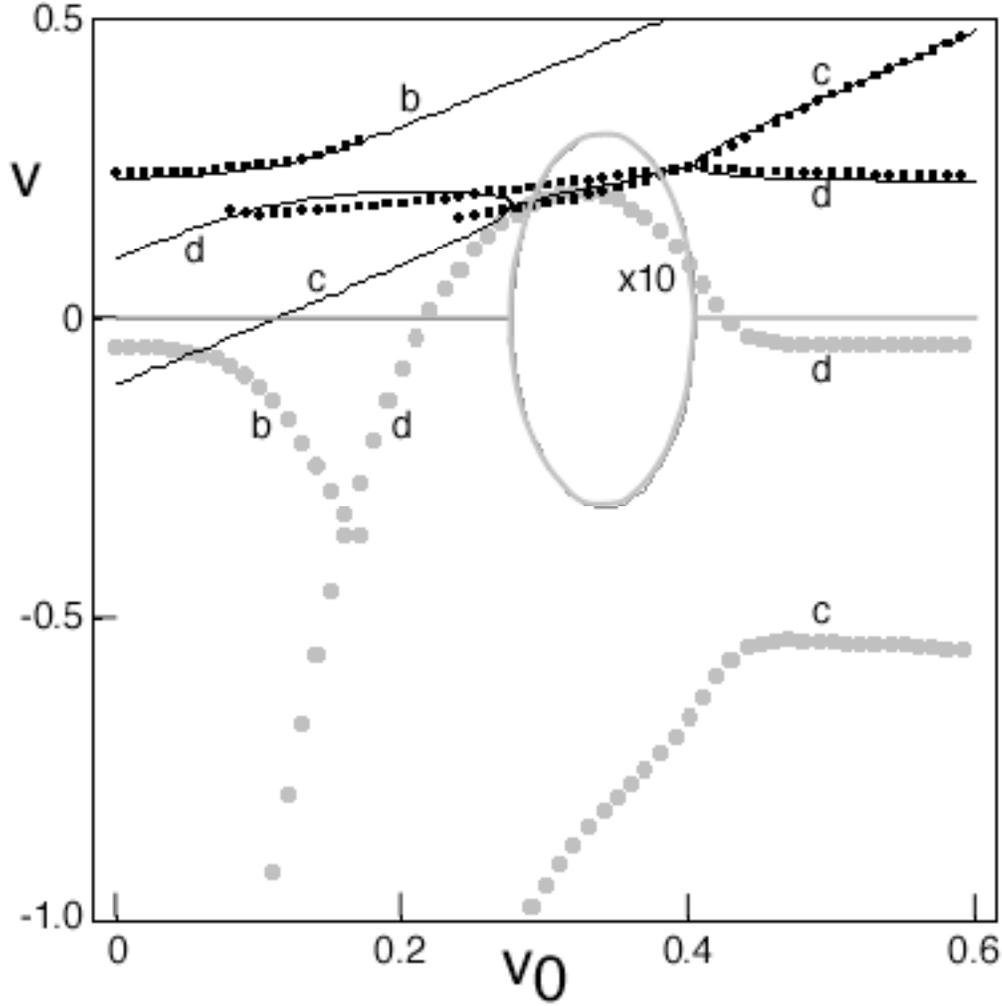

**Figure 5**. The smooth curves are the same as those in Figures 1 and 4 but for $T_e = 14 \times 10^6 K$, $T_p = 10^6 K$, $T_i = 4T_p$, $\gamma_e = 1$, $\gamma_{p,i} = 3$, $v_{Ap} = 1500$ km s$^{-1}$ and $N = 0.06$. The black points give the real part of $v$ at $\kappa_z = 0.01$ obtained via NHDS (with θ=80 degrees) for selected $v_0$, while the gray points give 10 times the corresponding imaginary parts of $v$. On the whole this validates our analysis, with the addition of weak damping from the Landau resonance.

In Figure 5, we show solutions to the fully kinetic dispersion relation for a hot plasma consisting of Maxwellian protons, electrons, and alpha particles. We obtain the solutions from the numerical solver NHDS (Verscharen et al., 2013). Acoustic modes in a linearized Vlasov-Maxwell system have phase speeds consistent with one-dimensional adiabatic ions ($\gamma_{p,i} = 3$) and isothermal electrons ($\gamma_e = 1$) (Gary, 1993). We use the same parameters as in Figure 4, but with these values of $\gamma$, and we increase the electron temperature to $T_e = 14T_p$ in order to reduce the Landau/transit-time damping to the point where a comparison with our analytical results can be made, even though we realize that these parameters do not correspond to the corona or solar wind. We set the angle θ between **k** and $\mathbf{B}_0$ to 80°. The smooth curves are our analytical solutions (though solution 'a' in Figures 1 and 4 has been omitted here), the black points are the real parts of $v$ obtained via NHDS for selected values of $v_0$, while the gray points are the imaginary parts, multiplied by 10 as in Figures 1 and 4. The kinetic solutions follow the analytical solutions well. The non-resonant instability is present, along with the cusp-like and ion-dominated roots, susbstantiating our analysis. However the instability appears at reduced growth rate due to the Landau/transit-time damping. As already noted, the instability appearing in our analysis is consistent with Equation (31).

The example in Figure 5 was chosen to give only weak damping. But by varying only $T_e$, we find that this instability is present only when $T_e / T_p > 7$, which still does not correspond to conditions in the corona or solar wind.

We emphasize that this kinetic calculation is based on the assumption of a Maxwellian distribution function for all species. If any distribution function is in a relaxed marginally stable state, its contribution to the damping can be lower or even negligible. The nonlinear treatment of the evolution of the plasma distribution functions and the associated wave-particle interactions, however, is beyond the scope of this work.

The NHDS solutions give also the electric field polarizations, and reveal that $\delta E_y$, which was ignored in the analysis of Section 2, can be comparable to or even larger than $\delta E_x$ or $\delta E_z$. Yet Figure 5 shows that our derived dispersion relation agrees well with the NHDS solutions for the real part of $v$. The reason is as follows: If we include $\delta E'_y$, we find that, consistent with our expansions in $\Omega^{-1}$, equations (8), (10) and (11) are unaltered. And our equations for $\delta V_x$, which were used to derive $\delta j_x$ and ultimately the dispersion relation, are then only modified by the addition of $c\,\delta E'_y / B_{0z}$. These $\delta\mathbf{E} \times \mathbf{B}_0$ drifts make no contribution to $\delta j_x$ or to the dispersion relation as long as the plasma is quasineutral and there is no background current $j_{0z}$.

The presence of $\delta E_y$ is easily understood. From $\nabla \times \delta \mathbf{E}_z$ Faraday's law implies the possibility that $\delta B_x \neq 0$. Ampere's law would then require $\delta j_y$, which in our low-frequency situation must arise from polarization drifts in the y-direction, which in turn require $\partial \delta E'_y / \partial t \neq 0$. A full discussion is beyond the intent of this paper, but it is useful to consider as an example the electron-proton plasma. Following the prescription which we just outlined (but with $\delta \mathbf{E}_y$ and $\delta \mathbf{E}_z$ included in $\nabla \times \delta \mathbf{E}$), we obtain

$$\frac{\delta E_y}{i \delta E_x} = \frac{\tan^2 \theta \, \gamma_e \, \beta'_e \, \nu \, \kappa_z}{2(1-\nu^2)(\nu^2 - \beta_{ep})} \qquad (32)$$

and

$$\frac{\delta E_y}{\delta E_z} = \frac{\tan \theta}{1-\nu^2} \qquad (33)$$

where $\theta$ is the angle between $\mathbf{k}$ and $\mathbf{B}_0$, $\beta'_e = 8\pi n_{0e} K T_e / B_0^2$, $\kappa_z = k_z v_{Ap} / \Omega_p$, and we have used equations (10) and (11) which give

$$\frac{\delta E_z}{i \delta E_x} = \frac{\tan \theta \, \gamma_e \, \beta'_e \, \nu \, \kappa_z}{2(\nu^2 - \beta_{ep})} \qquad (34)$$

With $\nu^2$ given by equation (19a), and with the parameters used in Figure 5, we find that equations (32) - (34) give good approximations to the NHDS polarizations of the cusp-like solutions, even those with $v_0 \neq 0$. The ratio $|\delta E_y / \delta E_x| < 1$ only for $\kappa_z < 8 \times 10^{-3}$ or so, and increases linearly with $\kappa_z$; the electric field components transverse to $\mathbf{B}_0$ are in general elliptically polarized and rotate in the left-hand (ion-resonant) sense. And for large $\theta$, $\delta E_y / \delta E_z > 1$ for all values of $\kappa_z$; in fact equation (33), which turns out to be an excellent approximation to the kinetic results, implies $(\nabla \times \delta \mathbf{E})_x \neq 0$ and thus $\delta B_x \neq 0$. Note that $\delta E_y$ can be significant even when $k_z^{-1}$ is much larger than the proton inertial length and $k_y^{-1}$ is much larger than the proton gyroradius; standard MHD could not have led to this conclusion.

## 5. Summary

We have studied the MHD slow mode in a plasma consisting of fully-ionized hydrogen and a heavy ion (taken to be fully-ionized helium in the numerical examples) drifting along the background magnetic field. Such a plasma is representative of the solar wind and the solar corona close to the Sun where the solar wind becomes supersonic. Using a three-fluid analysis, we presented a dispersion relation for both the fast and slow modes (equations 14 – 15b), but since the corona and solar wind contain many structures and fluctuations with large ratios of perpendicular to parallel wavenumbers we specialized to propagation highly oblique to the magnetic field (equation 18). This limit is also representative of the 'sausage mode' on thin coronal structures, which has been observed optically and has proven useful in coronal seismology. The slow mode can also be produced via parametric instabilities of the Alfvén waves which are copiously present in the solar wind, and the slow mode may also be associated with MHD turbulence in the solar wind. We have emphasized the fact that the highly oblique slow mode has very small variations of total pressure, and might be confused with non-propagating pressure-balanced structures or with the kinetic Alfvén wave; the results of this paper should be useful in distinguishing between these alternatives.

Figures 1 and 4 present the dispersion relations for two illustrative cases. For small ion concentrations we find two modes (forward and backward propagating) which are close to the usual 'cusp-like' modes in an electron-proton plasma (equation 19a). We also find two modes which are dominated by the convecting ion (equation 25). For reasonable coronal and solar wind parameters, these two types of modes exhibit an avoided crossing at small ion drift, and then cannot be simply characterized. Thus, when applied to coronal seismology, the standard dispersion relation for the sausage mode in an electron-proton plasma may lead to non-trivial errors.

We found that the drifting ions can drive a non-resonant instability, albeit only in a limited range of ion drift speeds $v_0$. For low plasma β's, the drift necessary for instability can be less than that required to drive electromagnetic ion-resonant modes (Verscharen & Chandran 2013). Although we have not carried out a quasilinear analysis, it is reasonable to suppose that the instability would react back on the ions in such a way as to limit their drift, and that the drift energy will be deposited in the plasma as heat. But by supplementing our analysis with fully-kinetic solutions for the dispersion relation we have found that with Maxwellian distribution functions the damping associated with the Landau resonance can eliminate the instability altogether, unless $T_e / T_p >> 1$. Our ideal fluid analysis also ignored damping associated with viscosity, heat conduction, and inter-particle collisions.

A comparison with solutions of the full dispersion relation of a hot plasma shows that the non-resonant instability of this study is the counterpart of the kinetic ion/ion-acoustic instability. Landau damping by the protons can efficiently counteract the instability driven by the ions if the particles have a Maxwellian distribution function and temperatures not greatly exceeding the proton temperature.

We have also found (equations 32-34) the surprising result, not contained in MHD, that $\delta E_y$ can be significant even when $k_z^{-1}$ is much larger than the proton inertial length and $k_y^{-1}$ is much larger than the proton gyroradius. But the dispersion relation is not affected by $\delta E_y$.

We found that the protons and ions do not move together (equation 24 and Figure 2), and that their compressions are not proportional, i.e. $(\delta n_i / n_{i0})/(\delta n_p / n_{p0}) \neq 1$ (equation 23 and Figure 3), even when the ions are not drifting. The velocity and density fluctuations of the two species can even be out of phase. One consequence is that even with zero drift, the standard dispersion relation (17) for the sausage mode cannot be used with $v_A$ computed using the summed proton and ion densities, and with $v_s$ computed from the summed pressures and densities of the electrons, protons, and ions. However, the relationships between the velocity and density fluctuations of the protons and ions may be used to identify the waves, and distinguish the oblique slow mode from non-propagating pressure-balanced structures. There is however a caveat: a fluid analysis (Hollweg 1999) suggests that the kinetic Alfvén wave also has $\delta p_{tot} \approx 0$, and may be confused with pressure-balanced structures and the highly oblique slow mode.

Finally, it has not escaped our notice that the drifting "ions" in our analysis can be taken to be the proton 'beams' which appear routinely in *in situ* observations of proton distribution functions in the nearly collisionless fast solar wind (see Figure 8.1 of Marsch 1991; Marsch 2012). This case will be derferred to a future publication.

**Acknowledgments:** This work has been supported by NASA grant NNX11AJ37G from NASA's Heliophysics Theory Program, NASA grant NNX13AF97G, and NSF grant AGS-1258998